\documentclass{jpsj-suppl}
\usepackage{txfonts} 
\title{Radio detection of cosmic rays: present and future}
\author{Tim \textsc{Huege}$^{1}$ and Andreas \textsc{Haungs}$^{1}$}

\inst{
$^{1}$Institut f\"ur Kernphysik, Karlsruhe Institute of Technology (KIT), Germany
}

\email{tim.huege@kit.edu, andreas.haungs@kit.edu}

\abst{Digital radio detection of cosmic rays has made tremendous progress
over the past decade. It has become increasingly clear where the potential
--- but also the limitations --- of the technique lie. In this article,
we discuss roads that could be followed in future radio detection efforts
and try to evaluate the associated prospects and challenges.}

\kword{UHECR 2014, radio detection, cosmic rays}

\begin{document}
\maketitle

\section{Introduction}

Over the past decade, tremendous progress has been made in the field of 
digital radio detection of cosmic ray air showers --- for an overview 
we kindly refer the reader to the short review and references 
presented in \cite{huegeicrcreview}.
Based on promising results achieved with the first-generation experiments
LOPES \cite{lopes_ldf} and CODALEMA \cite{torres2013}, a second generation of experiments
including in particular AERA \cite{aera}, LOFAR \cite{lofar} and 
Tunka-Rex \cite{tunka-rex} has
been set up with the goal to evaluate and exploit the true potential
of the detection technique.

In parallel to the experimental activities, the needed simulation 
codes and analysis strategies 
have been developed. Based on a solid understanding of the underlying 
radio emission physics, analyses have been devised to reliably 
extract the geometry, energy and even depth of shower maximum of air 
showers measured with radio antennas. In parallel to these 
achievements, however, some intrinsic limitations of the detection
of radio emission from air showers on large scales have become increasingly clear.

Based on the current state of the field, we try to evaluate where the 
most promising potential of radio detection of cosmic rays in future 
cosmic ray research lies. We stress that this is our personal view which
might not necessarily be shared by the whole community. Nevertheless, 
this contribution can certainly act as a good starting point for further discussions.

\section{The promises of radio detection}

Renewed interest in radio detection of cosmic rays at the beginning of 
the millennium had been sparked by several promising aspects of the 
technique. We give a concise overview here and try to summarize 
shortly to what extent these promises have indeed been fulfilled.

\begin{itemize}
\item Radio emission gives information complementary to particle 
detectors: This is true in the sense that radio detection purely 
measures the electromagnetic component of an air shower, while 
particle detectors can be sensitive to various components of the 
cascade. Complementarity can be maximized in particular if particle 
detectors specifically measure the muonic component of air showers.
\item Radio measurements provide a calorimetric energy measurement: 
The radio signal at MHz frequencies is coherent and thus its amplitude 
scales linearly with the number of particles, which is in turn roughly 
proportional to the energy of the primary particle. As there is no 
absorption of the radio signal in the atmosphere, indeed the radio 
signal provides a calorimetric measurement of energy in the electromagnetic 
component of an air shower \cite{lopes_ldf}. Furthermore, energy estimators can be chosen 
which are hardly influenced by shower-to-shower fluctuations. Intrinsically (without accounting for experimental uncertainties) the energy resolution achievable with radio detectors should be better 
than 10\%, potentially even as good as 5\% \cite{lopes_ldf,huegeulrichengel2008}.
\item The depth of the air shower maximum (Xmax) can be measured with the radio 
technique: Due to the forward-beamed nature of the radio emission, the 
distance of the radio source, closely correlated with the depth of 
the shower maximum, leaves an imprint in several observables, 
in particular the slope of the radio lateral distribution function 
\cite{lopes_ldf,huegeulrichengel2008,lofar_xmax}, the opening angle of 
the hyperbolical wavefront \cite{lopes_wavefront} and 
the pulse shape. Xmax resolutions of 
20~g/cm$^{2}$ or better can in principle be achieved, although the 
experimental confirmation with independent cross-checks using, e.g., the
fluorescence detectors of the Pierre Auger Observatory or the Cherenkov-light
detectors of Tunka has yet to be performed.
\item Radio has a near-100\% duty cycle: It has been established that 
radio emission is strongly influenced during thunderstorm conditions 
\cite{lopes_thunderstorms}. In fair weather conditions, however, measurements can be made 
day and night. Depending on the location, duty cycles of $\approx 
95$\% seem to be achieved. This high duty cycle is coupled with a field-of-view of 
nearly $2 \pi$, which is considerably larger than the one usually 
achieved with particle, fluorescence light and Cherenkov light 
detectors. (It should be noted, though, that the radio detection 
threshold depends significantly on the arrival direction.)
\item A high angular direction resolution is achievable: Even small 
arrays such as LOPES have demonstrated an angular resolution below 
1$^{\circ}$. The angular resolution is expected to be better than 
0.5$^{\circ}$ for more sophisticated and larger arrays.
\item The detectors can be built cheaply: This depends on two main 
aspects, the cost of an individual radio detector element and the 
needed density of these elements. Of the current experiments, only 
Tunka-Rex has been designed with the explicit goal of cost-effectiveness. 
Tunka-Rex has been built with a cost of only 500 USD per 
antenna \cite{tunka-rex} --- however, significant infrastructure of the existing Tunka 
array was used in the process. It is clear that antennas and digital 
electronics can certainly be built at much lower prices than those of 
particle detectors. The difficulty for a cheap radio detector arises, 
however, when the needed spacing of the detectors is considered, which we will 
discuss below.
\end{itemize}

We will now discuss the potential for various specific implementations 
of radio detection of cosmic rays in more detail.

\section{Potential of sparse arrays}

Radio detection of cosmic rays in the frequency range of $\approx$~30 
-- 80~MHz has started with small-scale radio detection setups covering 
areas of less than 0.1~km$^{2}$. With those it was established that 
the detection threshold for radio emission lies approximately at 
$10^{17}$~eV, where the radio signal starts to dominate over the 
Galactic Noise. Although the details depend on the local magnetic 
field configuration, the arrival direction, the altitude of the 
experiment and the analysis technique (interferometry can improve the 
signal-to-noise ratio), this threshold can be considered an adequate rule of thumb.

From the beginning, the interest in radio detection was to extend the 
energy reach of the measurements, ideally to be able to measure at the 
highest energies up to $10^{20}$~eV. As the cosmic ray flux drops very 
dramatically with primary particle energy, this requires the 
instrumentation of very large areas. To keep instrumentation cost-effective, 
radio antennas have to be spaced as far apart as possible while still 
ensuring efficient detection. Hence, one of the goals was to make arrays as 
``sparse'' as possible.

\begin{figure}[tbh]
\centering
\includegraphics[width=0.85\linewidth]{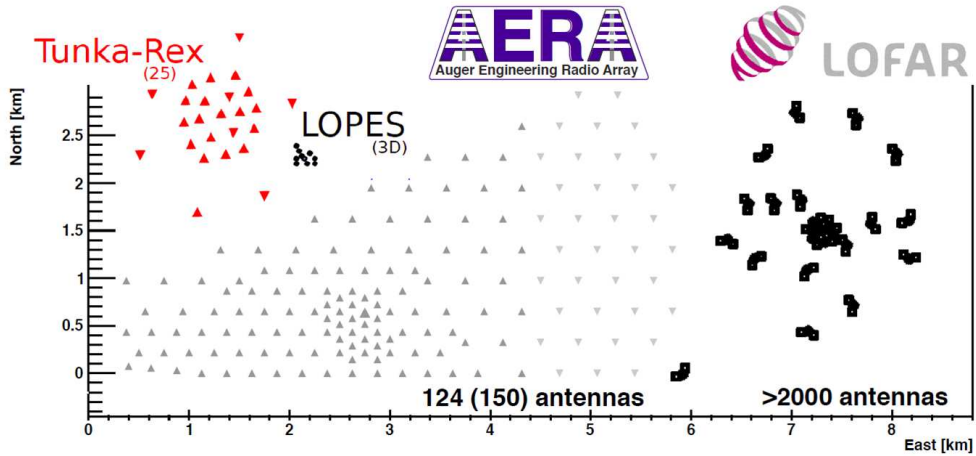}
\caption{Comparison between the layouts of LOPES, Tunka-Rex, AERA and LOFAR. 
While Tunka-Rex and AERA try to cover large areas with few antennas 
(sparse arrays), LOFAR consists of ``stations'' with hundreds of 
antennas on very small areas. The large gaps between these dense areas 
are unfavourable for air shower detection but suit interferometric 
observations of astronomical objects. The Diagram was compiled by D.\ Huber, 
F.G.\ Schr\"oder and J.\ H\"orandel.}
\label{fig:experiments}
\end{figure}

\begin{figure}[tbh]
\centering
\includegraphics[width=0.7\linewidth]{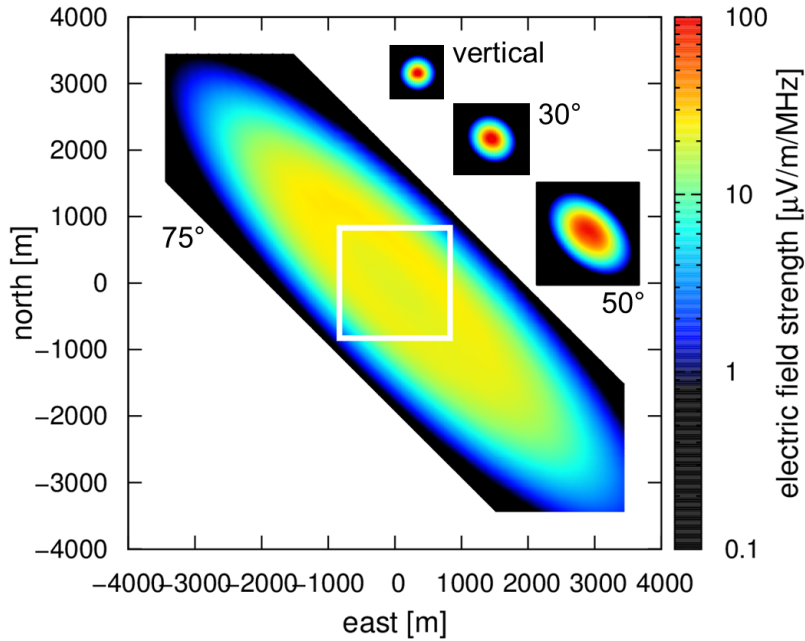}
\caption{Radio emission footprints in the 30-80~MHz frequency band as 
simulated with CoREAS \cite{coreas}. The detection threshold typically 
lies at $\approx$~1-2~$\mu$V/m/MHz. While the radio emission footprint 
is small for air showers with zenith angles up to $\approx 
60^{\circ}$, it becomes very large for inclined showers with zenith 
angles of $70^{\circ}$ or larger. The white rectangle characterizes 
the size of the $50^{\circ}$ inset. The dramatic increase of the 
illuminated area with the shower zenith angle is caused by the increasing
geometrical distance between the shower maximum and the ground.}
\label{fig:inclined}
\end{figure}

The Auger Engineering Radio Array (AERA) has pursued this approach 
with a graded array layout, which can be seen in comparison with LOPES-3D, 
Tunka-Rex and LOFAR in Fig.\ \ref{fig:experiments}. The spacing of 
the AERA stations has been set to grids of 144~m, 250~m and 375~m and 
in the last phase 750~m. It 
has become increasingly clear that the size of the ``radio footprint'' 
illuminated by an air shower is dominated completely by one parameter: 
the zenith angle of the air shower. For near-vertical showers the 
footprint is very small due to the closeness of Xmax to the ground. A 
spacing of 375~m is probably already too sparse for coincident 
detection of near-vertical 
showers with at least three antennas. When more inclined geometries are considered, the picture 
changes dramatically, as is illustrated in Fig.\ \ref{fig:inclined}. 
For the detection of showers with zenith angles of 70$^{\circ}$ or 
more, a grid of 750~m or even 1.5~km (the baseline grid of the Pierre 
Auger Observatory) can be sufficient.

A very important fact to realize is that the energy of the primary 
particle will not mitigate the effect of the small footprints for 
near-vertical air showers. While the electric field amplitudes do grow 
linearly with primary particle energy, the steepness of the lateral 
distribution function still limits the extent of the radio emission 
footprint to small areas.

Two consequences arise from these simple considerations (which, in 
fact have already been hinted at in the review article of Allan in 
1971 \cite{Allan1971}): A radio detection array focused at air showers with zenith angles 
below $\approx 60^{\circ}$ zenith angle will need to be fairly dense if 
coincident detection in three or more antennas is foreseen. A spacing 
of $\approx 300$~m seems to be adequate. If the goal is to cover large 
areas, this requires very many antennas to be deployed. To reach the 
same yearly exposure as the Auger Fluorescence detector, an area of $\approx 
300$~km$^{2}$ would need to be covered (a tenth of the area with 
roughly ten-fold duty cycle). At a spacing of $\approx$~300~m this requires of 
order 3000 -- 5000 antennas. This would be a tremendous effort. A very 
significant cutting of the 
cost of individual antenna stations with respect to those of 
current-generation detectors such as in AERA is certainly possible. Consider that
a modern-day smartphone has many of the 
capabilities needed by a radio detector, and that by Moore's law, the 
needed digital electronics drops in price very significantly year by 
year. Wireless communication of the acquired data could profit from 
new developments like LTE-Advanced in which the capability to form ad-hoc mesh networks
receives special attention. Communications thus seems a solvable problem. The biggest challenge in fact seems to 
be power harvesting: solar power is expensive and requires expensive 
buffering with batteries. Maybe cheap wind turbines could be an option 
if they can be made reliable and radio-quiet. However, even if the cost 
per detector station is reduced to a few hundred USD, another problem would be the manpower needed for deployment and maintenance, 
which would have to be kept to an absolute minimum. This seems to call 
for a visionary (or 
crazy?) approach such as "deploy by airplane", i.e. pour a 
huge number of cheap autonomous radio detectors out of an airplane or 
helicopter while flying over the area to be instrumented. 
While none of this seems impossible per se, it seems clear that the 
concepts developed so far can not simply be scaled to areas much 
bigger than a few dozens of km$^{2}$ area.

This, however, is certainly not to say that radio detection on the scales achieved today 
cannot make important contributions to cosmic ray physics. Very 
strong potential exists in independent cross-checks of the energy scale 
of existing particle and fluorescence detectors. This is because radio 
emission can be predicted from first principles on the basis of the 
pure electromagnetic cascade of an air shower. There are no 
uncertainties such as an unknown ``yield'' and there is no absorption 
in the atmosphere. The biggest challenge is a precise absolute 
calibration of the radio detectors, in particular the antenna 
response. Once this is available, radio detection can be used as a 
very precise technique to independently measure the absolute energy 
scale of cosmic ray measurements and to precisely determine the energy on an 
event-to-event basis. Also, Xmax reconstruction on an event-to-event 
basis with arrays such as AERA seems to be 
promising, although the actual resolution still has to be established 
experimentally.

Another direction that seems promising to pursue is the detection of 
inclined air showers with zenith angles of 70$^{\circ}$ or more. Here, 
sparse radio detection arrays with grid spacings of a km or more 
should suffice. As these scales are comparable to those on which 
particle detectors are deployed, an integration of radio detectors 
with particle detectors seems very attractive --- this allows re-use of 
much of the infrastructure for power harvesting and wireless 
communications, making the additional cost rather small. The benefit 
would be that while particle detectors for such inclined air showers 
basically measure pure muons (the electromagnetic component has died 
out when the shower reaches the ground), radio detection would provide 
a precise measurement of the electromagnetic component of the air 
shower. Combined, this can be a powerful tool for mass composition 
and/or air shower physics studies, in which not only the energy scale 
but also further details could be probed.

Finally, one should not forget another option, which is to make use of 
the information provided by the radio detection of an air shower with 
only one antenna. If the geometry of the air shower (in particular the 
core position) is known precisely from another detector such as a particle 
detector array, even a single radio antenna measurement can provide a 
wealth of additional information. The best option for an optimal local 
hybrid detector as well as the analysis techniques for this 
approach have yet to be developed, though. Of course, the probability to detect
an air shower with even one antenna still depends strongly on the density of detectors,
especially for near-vertical showers.

\section{Potential of dense arrays}

A complementary approach to trying to instrument the largest possible 
area with the smallest possible number of antennas is to instrument a 
given area very densely with radio antennas. The Low Frequency Array 
(LOFAR) can be considered such a dense array. In particular, in its 
dense core hundreds of antennas are spaced in an area with a diameter 
of only 400~m. The layout of the LOFAR antennas inside and outside 
this core was optimized for interferometric observations of 
astronomical radio sources and is not 
ideal for cosmic ray radio detection, as there are very densely 
instrumented ``stations'' with large gaps in between. Nevertheless, 
LOFAR has shown very impressively how much detail can be extracted 
from radio measurements of air showers with a dense antenna array. In 
particular, an Xmax resolution of better than 20~g/cm$^{2}$ has been 
achieved with room for further optimisation of the analysis technique 
\cite{lofar_xmax}. (It should be noted, though, that this Xmax reconstruction
so far purely rests on simulations of the radio emission from extensive air showers.
LOFAR lacks the capability for independent experimental cross-check of this result as it does not 
have access to Xmax information measured with another detection 
technique. This experimental verification will have to be achieved with AERA and Tunka-Rex in the near 
future.)

\begin{figure}
\centering
\includegraphics[clip=true,trim=80 28 90 35, width=0.32\textwidth]{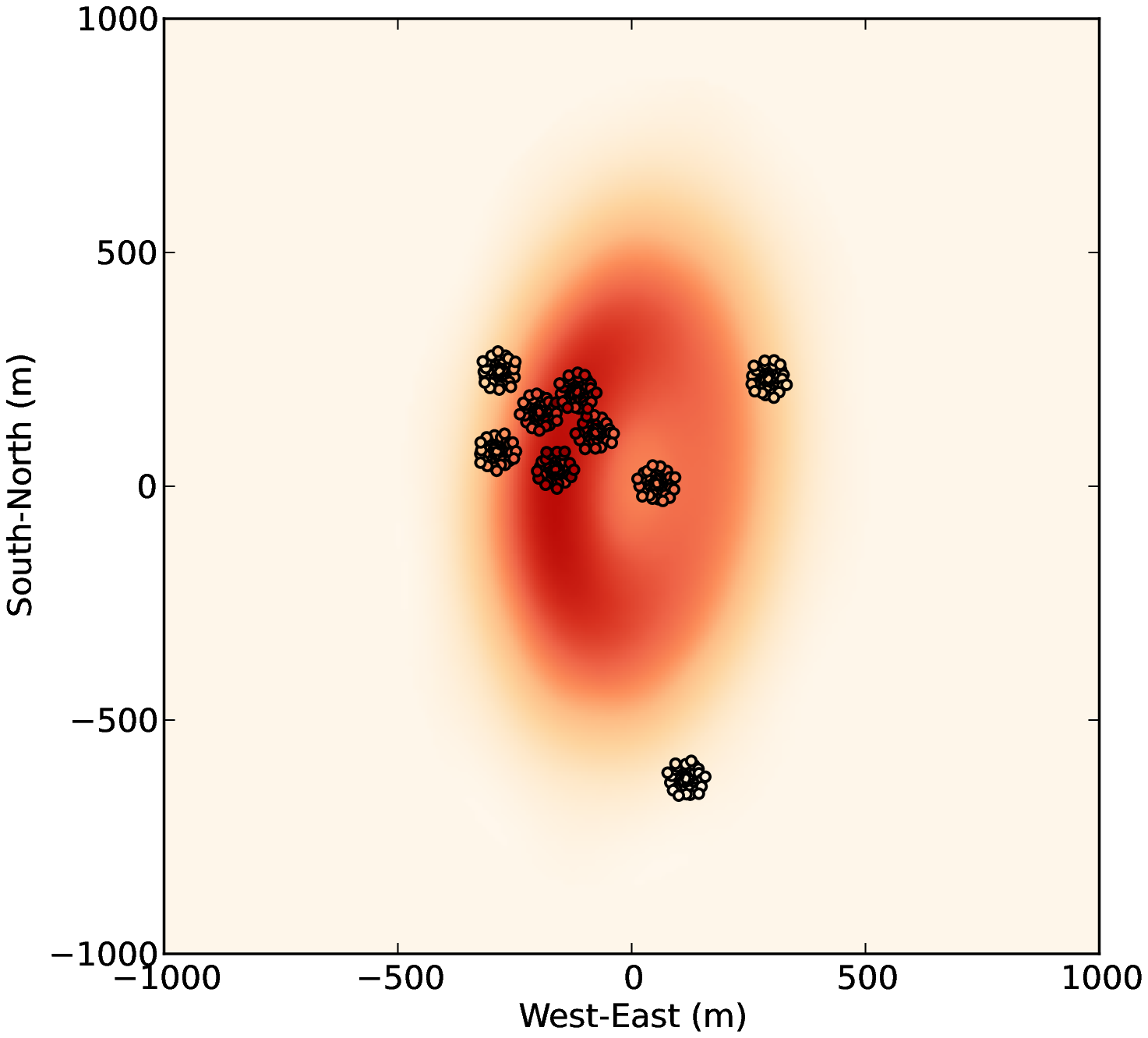}
\includegraphics[clip=true,trim=80 28 90 35, width=0.32\textwidth]{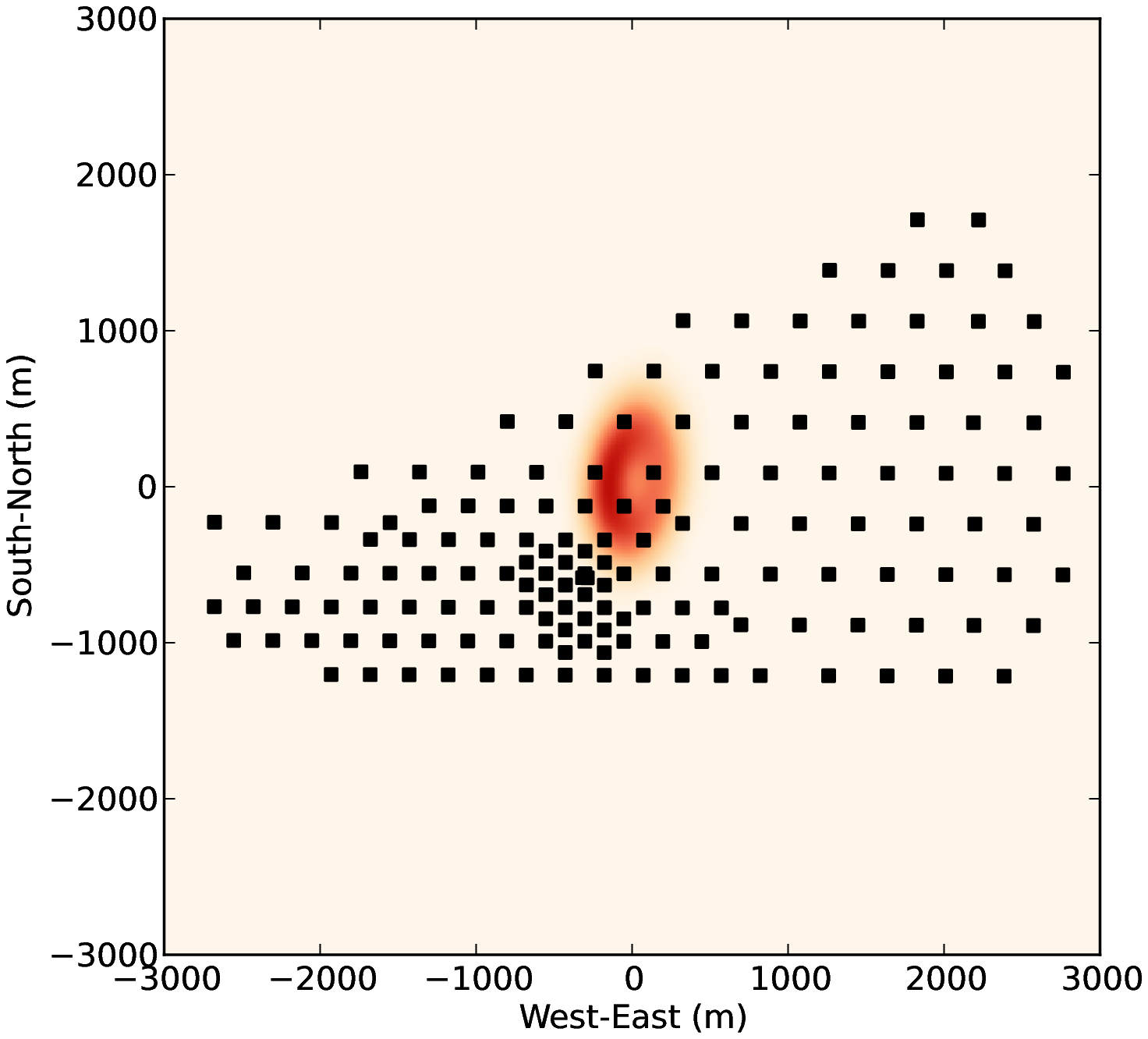}
\includegraphics[clip=true,trim=80 28 90 35, width=0.32\textwidth]{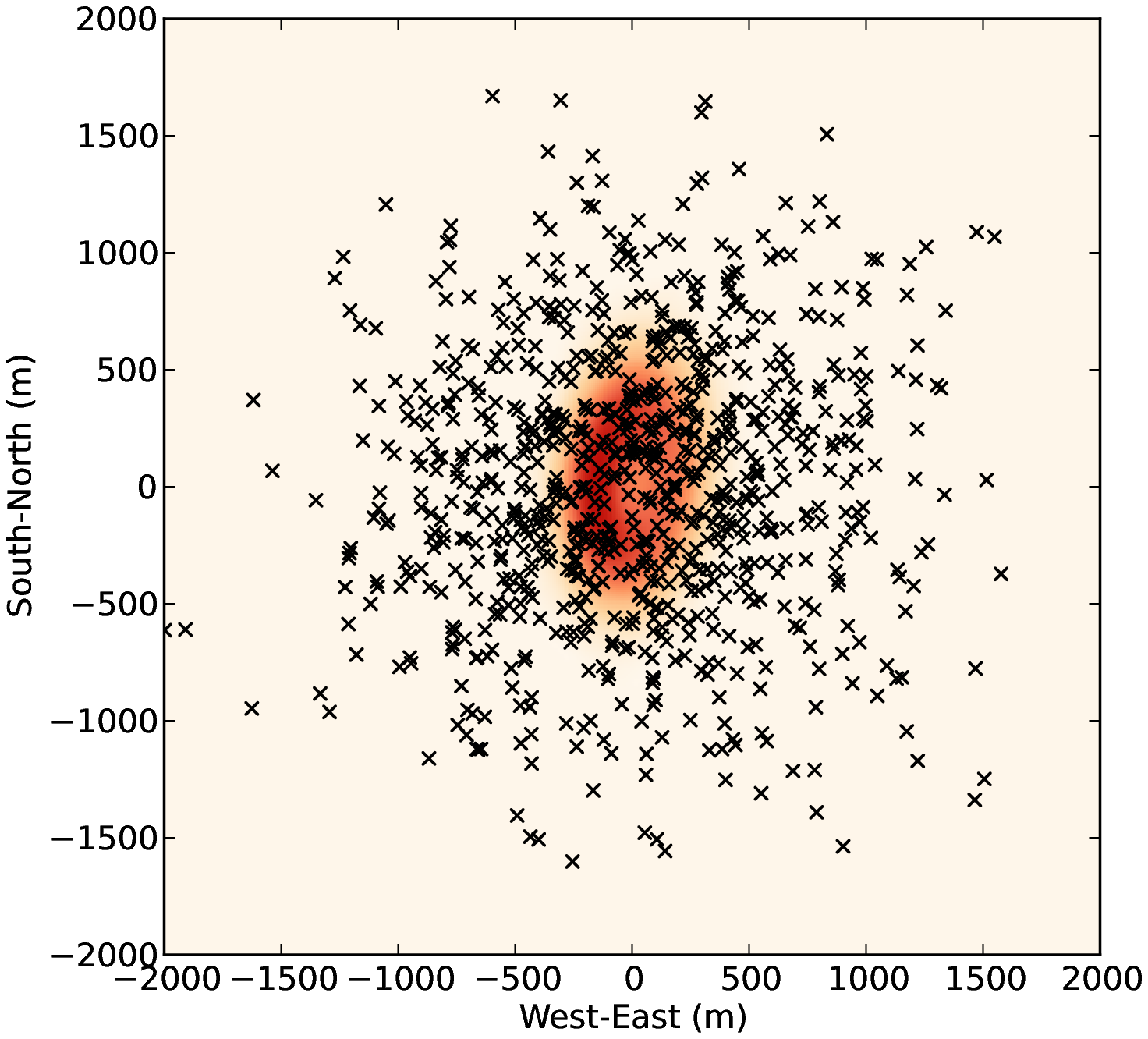}
\caption{Antenna layouts for LOFAR (left), AERA (middle) and a 
a simulation of the core region of the SKA (right) (axes denote 
distances in metres). The background colors represent the radio 
emission footprint of an air shower with a zenith angle of 
55$^\circ$ as simulated with CoREAS \cite{coreas}. Both LOFAR and SKA 
sample the footprint with hundreds of antennas simultaneously, but at 
SKA the coverage will be much more uniform. Diagrams are from 
\cite{ska-eas}.}
\label{fig:detector_layouts}
\end{figure}

While the concept of such dense arrays does not scale to the large 
detection areas needed for the study of ultra-high-energy cosmic rays, 
there is strong potential for ``precision studies'' of air shower 
and cosmic ray physics at energies from $10^{17}$~eV to 
$\approx$~10$^{19}$~eV, i.e., in the region of transition from Galactic to 
extragalactic cosmic rays. In particular, the low-frequency part of 
the coming Square Kilometre 
Array (SKA) will instrument a significantly larger area than LOFAR 
with a very dense homogeneous array of radio antennas sensitive in the 
frequency range from 50 to 350~MHz, see Fig.\ 
\ref{fig:detector_layouts}. Activities are ongoing to ensure 
the needed particle triggering and buffering capabilities to use SKA-Low as an 
air shower detector in parallel to astronomical observations 
\cite{ska-eas}. With the 
SKA, precision studies of the mass composition in the transition 
region, with Xmax resolutions on the level of up to $\approx$~10~g/cm$^{2}$ could be 
pursued. Also, the SKA could be used to study particle interactions at 
the highest energies by precise measurements of the electromagnetic 
component of extensive air showers. Using (yet-to-be-developed) 
near-field interferometric analysis techniques it should be possible to 
reconstruct a three-dimensional ``tomography'' of the electromagnetic 
component of the air shower, offering a wealth of information for the 
study of air shower and thus particle physics. Considering the 
relatively small design changes needed to use the SKA for such studies 
and the potential scientific gain it is clear that this is a highly 
attractive option for the future application of radio detection of air 
showers.

A way with which dense arrays can be used for the observation of cosmic 
rays up to the highest energies is to not use the atmosphere as 
interaction medium but rather use the moon as active volume. In this 
scheme, which has been followed for decades with various projects, but 
without successful detection to date, several ``beams'' are placed 
simultaneously on the lunar limb to search for Askaryan-effect radio pulses arising 
from particle cascades initiated in the lunar regolith. The advantage 
of using the moon is that the detection volume is tremendous. The 
disadvantage is that the source is very far away and thus the radio pulses 
become very weak, requiring a very sensitive radio telescope to pick 
them out. With the SKA, for the first time detection of cosmic rays at 
energies below 10$^{20}$~eV should become feasible. Such observations 
would allow pin-pointing of the sources of cosmic rays, and there is 
even a moderate energy resolution expected \cite{ska-lunar}. Consequently, this is another 
way of cosmic ray detection using dense radio arrays which is being 
followed very actively. At the same time, this technique can also be used to search for 
ultra-high energy neutrinos, exploiting the vast detection volume 
available for neutrino interactions in the moon. 

\section{Potential of radio-only detectors}

From the start of digital radio detection activities, one prominent question 
was how a stand-alone radio detector could be realized. Significant 
effort has been made to develop a ``self-triggered radio detector'' 
and some successes have been achieved \cite{rauger}. In principle, it is of 
course possible to trigger read-out of radio detectors on the basis of 
pulses measured in the radio signals themselves. However, it has 
turned out to be very challenging indeed to reliably identify cosmic ray air 
showers among the triggered signals. The problem is that there are 
ample sources of pulsed radio signals, in particular of anthropogenic 
origin. This reaches from faulty isolation in power lines or 
transformers kilometres 
away over airplanes passing above the detector to lightning in 
thunderclouds many kilometres away from the detector. The rate of such 
``transient RFI'' is what limits self-triggering capabilities, while 
the ``continuous noise'', in particular the Galactic noise, governs 
the overall detection threshold. It is thus not sufficient to survey a 
location by taking a spectrum analyzer and having a look at the 
frequency spectrum of the measured noise, as has sometimes been done 
in the past. The one site that remains to seem particularly promising 
for self-triggering of radio detectors is the south pole, which seems 
to be virtually free of transient RFI at MHz frequencies according to 
measurements by the RASTA project \cite{rasta}.

A self-triggered radio detector is certainly not impossible. Some
successes have been achieved, and the radio signal exhibits enough 
peculiarities that should allow identification of air showers even in the 
presence of anthropogenic RFI (the lateral distribution of the radio 
signals over the antennas, the hyperbolic wavefront, the peculiar 
polarization). However, the power of radio detection does not lie in 
isolated measurements without any other detector. The true potential 
lies in combining different detectors in the most complementary way, 
for example using muon detectors in combination with radio antennas. 
Not only does this provide a completely different quality of the 
measurement, it is also attractive from a cost point of view since 
much of the infrastructure can be shared by the different detectors. 
It is so much easier to then use the trigger provided by the particle 
detectors, which exhibits virtually no false positives and 
also is able to achieve a significantly lower trigger threshold than 
is possible on the basis of the radio signals alone. In other words, 
while a self-triggered radio detector is not impossible to build, this is so much 
harder to achieve than triggering by particle detectors, and there is 
so little convincing reason to do so that the question seems to have become 
purely academic today.

\section{Potential of high- and low-frequency observations}

There were great hopes to use radio detection at GHz frequencies for 
``fluorescence-like'' detection of air showers 24 hours a day with 
off-the-shelf hardware. This would be possible if there were 
isotropic radio emission at GHz frequencies by the ``molecular 
bremsstrahlung'' effect that was originally invoked to explain 
measurements at the Argonne Wakefield Accelerator and SLAC \cite{gorhamslac}. Several experiments have by now attempted to 
measure this isotropic microwave emission, but to no avail. If it 
exists, it is much weaker than originally expected, and probably too 
weak to be exploited in practice. The most 
convincing detection of GHz radio emission has been made by the CROME 
experiment \cite{crome}, but these results are fully consistent with forward-beamed radio 
emission arising from the very same geomagnetic and charge excess 
radiation seen at MHz frequencies undergoing a Cherenkov-like time 
compression due to the refractive index gradient of the atmosphere. 
The special condition to be able to observe this emission is that the 
antennas need to be located on the ``Cherenkov ring'' that is 
illuminated by high-frequency emission on the ground. While the GHz 
emission provides interesting information (e.g., the diameter of the 
ring provides information on the distance of the source and thus 
Xmax), this is a very constrained geometry for detection which requires 
an even denser array than MHz observations. From this point of view, 
high frequency emission alone does not seem an attractive scheme. If 
low-frequency detectors can easily be extended to higher frequencies than 
the 80~MHz typically used today, this could however be attractive.

While the radio emission from air showers extends to frequencies below 
30~MHz, detection at such low frequencies suffers from the 
presence of very strong atmospheric noise. It could be feasible to 
exploit a frequency window at $\approx $ 1--10 MHz which is relatively quiet 
during day because of absorption effects in the ionosphere. At these 
frequencies, one could potentially also observe radio emission from 
another mechanism, the near-instantaneous deceleration of charged shower particles 
entering the ground. This effect has been predicted by simulations and 
has been dubbed ``sudden-death radiation'' 
\cite{suddendeath}. While it is clear that the sudden stopping of 
charged particles will cause a radio pulse, it is, however, currently 
rather unclear how much of the emission exits the ground-air boundary and if indeed 
it will be observable by antennas situated directly on the ground. 
Experimental efforts to test this detection scheme are ongoing in the EXTASIS project.

\section{Potential of balloon-/satellite-based observations}

Another option that was and still is being discussed in various 
scenarios is to use radio detectors high up in the atmosphere to 
identify air showers in large volumes of the atmosphere by their radio 
pulses. ANITA has already achieved this by detecting both direct 
(up-going) air showers and radio emission from air showers that has 
been reflected off the antarctic ice \cite{hoover2010}.

However, recent calculations \cite{motloch} indicate that the strong 
forward-beaming of the radio emission strongly limits the achievable 
apertures. Even with satellite-based observations, apertures of at most $\sim 20,000$~km$^{2}$~sr
can be achieved, which means that they will not be able to 
surpass the existing large ground-based detector arrays such as the 
Pierre Auger Observatory. This is true even at MHz frequencies where 
the beam pattern is much broader than the ring-shaped beam pattern at 
hundreds of MHz observed by the ANITA balloon. Considering the efforts involved in realizing a
satellite-based experiment and the relatively short times over which balloons can 
operate, these approaches do not seem to be as attractive as was 
previously hoped.

\section{Conclusions}

Radio detection of cosmic ray air showers has made tremendous progress 
over the past decade. It can contribute in many important ways to 
cosmic ray research, in particular by independent measurements of the 
energy scale and the mass composition of cosmic rays. Limitations 
exist in particular in the size of the radio emission footprint which 
requires antennas to be spaced on grids not larger than $\approx 
300$~m --- except if one concentrates on inclined air showers or only 
strives to complement particle detector measurements with radio 
measurements in single antennas. Very dense arrays of radio detectors, 
in particular the upcoming SKA, on the other hand, will be able to do 
high precision studies of air shower physics and the mass composition 
in the transition region. Radio detection may thus not fulfill all 
hopes that were originally expressed, but it certainly has the 
potential to provide very valuable additional information in hybrid 
detector concepts within existing and future cosmic ray observatories.

\section*{Acknowledgements}

We would like to thank all our colleagues working in the field for 
a decade of inspiring collaboration in a truly pioneering spirit.

%\appendix
%\section{}
% Use the \verb|\appendix| command if you need an appendix(es). 
% The \verb \section| command should follow even though there is
% no title for the appendix(see above in the source of this file).

\end{document}